\documentstyle[12pt,epsf]{article}
\newcommand{\newc}{\newcommand}
\newc{\gsim}{\lower.7ex\hbox{$\;\stackrel{\textstyle>}{\sim}\;$}}
\newc{\lsim}{\lower.7ex\hbox{$\;\stackrel{\textstyle<}{\sim}\;$}}
\newc{\invpb}{\,{\rm pb}^{-1}}
\newc{\tanb}{\tan\beta}
\newc{\sgnmu}{{\rm sgn}\,\mu}
\newc{\gev}{\,{\rm GeV}}
\newc{\tev}{\,{\rm TeV}}
\newc{\fb}{\,{\rm fb}}
\newc{\neut}{\widetilde{N}}
\renewcommand{\char}{\widetilde{W}}
\newc{\bino}{\widetilde{B}}
\newc{\wino}{\widetilde{W}}
\newc{\higgsino}{\widetilde{H}}
\newc{\gravitino}{\widetilde{G}}
\newc{\slepl}{\widetilde\ell_L}
\newc{\slepr}{\widetilde\ell_R}
\newc{\sneut}{\widetilde\nu}
\newc{\misset}{/\!\!\!\!E_\perp}
\newc{\eg}{{\sl e.g.}}
\newc{\ie}{{\sl i.e.}}
\newc{\etal}{{\sl et al}}
\newc{\ord}{{\cal O}}
\newc{\mw}{M_W}
\newc{\mz}{M_Z}
\newc{\staul}{\widetilde\tau_L}
\newc{\staur}{\widetilde\tau_R}
\def\NPB#1#2#3{Nucl. Phys. {\bf B#1} (19#2) #3}

\def\PRD#1#2#3{Phys. Rev. {\bf D#1} (19#2) #3}
\def\PRL#1#2#3{Phys. Rev. Lett. {\bf#1} (19#2) #3}
\def\PRT#1#2#3{Phys. Rep. {\bf#1} (19#2) #3}

\def\ZPC#1#2#3{Zeit. f\"ur Physik {\bf C#1} (19#2) #3}

\def\RMP#1#2#3{{Rev. Mod. Phys. } {\bf #1} (19#2) #3}

\catcode`@=11
\long\def\@caption#1[#2]#3{\par\addcontentsline{\csname
  ext@#1\endcsname}{#1}{\protect\numberline{\csname
  the#1\endcsname}{\ignorespaces #2}}\begingroup
    \small
    \@parboxrestore
    \@makecaption{\csname fnum@#1\endcsname}{\ignorespaces #3}\par
  \endgroup}
\catcode`@=12

\begin{document}


{\hbox to\hsize{May 1996 \hfill IASSNS-HEP 96/55}}\par
{\hbox to\hsize{hep-ph/9605408}\par
\begin{center}
{\LARGE \bf Experimental Consequences of a \\ Minimal Messenger \\
Model for Supersymmetry Breaking\footnote{Work supported in part by the 
Department of Energy contract No.~DE-FG02-90ER40542 and by the 
Monell Foundation.}\\} 
\vskip 0.3in
{\bf K.S. Babu, Chris Kolda, and Frank Wilczek}\footnote{E-mail:
babu@sns.ias.edu, kolda@sns.ias.edu, wilczek@sns.ias.edu}\\[.05in]
{\sl School of Natural Sciences\\
Institute for Advanced Study\\
Princeton, NJ 08540}\\[.15in]
\end{center}

\begin{abstract}

We calculate the low-lying spectrum of new particles expected in a
minimal model wherein supersymmetry breaking at $\lsim 100\tev$ 
is indirectly transmitted to the Standard Model.   We calculate the
couplings of these particles relevant to their most nearly accessible
experimental signatures, and estimate those signatures quantitatively.
Running of various couplings plays a crucial role in the
phenomenology, specifically in generating an adequate $\neut_1 -
\slepr$ splitting.

\end{abstract}
 
\section{Introduction}

Since the widespread realization that successful unification of gauge
couplings can be achieved in models including the minimal
supersymmetric extension of the standard model (perhaps augmented
with complete
$SU(5)$ multiplets) \cite{unification}, 
anticipation of the imminent discovery of direct
evidence
for real -- 
as opposed to virtual -- supersymmetry has become almost palpable.
On the other hand, there has been no consensus regarding the 
form this discovery might take.  Theoretical control over the many
new parameters that arise even in minimal supersymmetric extensions of
the standard model is not adequate to guide one through a bewildering
number of choices that substantially affect the predictions for
observable phenomena.  

Recently a class of models that potentially enjoy great
predictive power has been identified \cite{models}.  
The leading idea of these
models is that the fundamental breaking of supersymmetry occurs in a
``hidden sector'' and is characterized by a scale 
$\Lambda \lsim 100\tev$.
This breaking is supposed to be conveyed to our observable
sector, including the minimal supersymmetric standard model, by a
specific, highly symmetric interaction term (to be discussed more
precisely below).   This is to be contrasted with the situation in a
class of models, commonly referred to as  
supergravity models, that has been much more popular.  In supergravity
models  
the fundamental symmetry breaking scale is 
$\Lambda \sim 10^9\tev$, and the hidden sector communicates with
the observable sector by gravitational-strength interactions.  A
decisive phenomenological difference between these classes of models
arises because of the different role played by the 
gravitino $\widetilde G$, which is 
essentially the Nambu-Goldstone fermion of
spontaneous supersymmetry breaking.
Its
mass is generically given as 
$M_{\widetilde G} = {\Lambda^2\over M_{\rm Planck}}$, whereas its couplings
to ordinary matter scale as $1\over \Lambda^2$ \cite{fayet}.  
In most supergravity
models the gravitino is massive and very weakly coupled, and 
although it could be significant cosmologically, it plays no very
direct role in ordinary phenomenology.  In the (relatively)
low energy messenger models the gravitino is exceedingly light, and
although its couplings are very feeble they can be such as to cause
the next lightest, photino-like particle $\widetilde {N_1} $ to decay via
$\widetilde {N_1} \rightarrow \gamma + \widetilde G$ within a distance that
might be difficult to resolve experimentally \cite{signals}.  
Generically, then, in
these models one will find that pair-production of supersymmetric
(R-odd) particles will lead eventually, perhaps after a cascade of
decays, to final states containing $\gamma \gamma$ and missing energy.
The recent observation \cite{cdf} of a dramatic event with a final state
containing hard  $e^+e^-\gamma \gamma$ and missing transverse energy by
the CDF collaboration therefore 
lends special interest to these models.

In this paper we shall calculate the low-lying spectrum and
interactions in a minimal messenger model of the kind mentioned, in
sufficient detail to estimate the most nearly accessible processes.
We find a remarkably tight and specific pattern of consequences.
In the course of the analysis we shall find that some apparent
difficulties  
of the model are resolved by careful attention to the
running of various couplings.  Specifically,  the tiny $\widetilde {N_1} -
\widetilde {e_R}$ mass splitting one finds using bare couplings, and which
if correct would render the interpretation of the observed 
$e^+e^-\gamma \gamma$ event problematic~\cite{dtw}, is found to be greatly
enhanced by their running (and by the inclusion of D-term effects).  
Similarly a possible difficulty with the apparent smallness of the $B\mu$
parameter, which if valid would imply the existence of a weak scale
axion, is automatically repaired.  


\section{Spectrum}
In the minimal messenger model (MMM) supersymmetry breaking is conveyed to
the standard model particles by the matter fields 
$(q',\ell')+(\overline{q}',\overline{\ell}'$) having the
quantum numbers ${\bf 5}+{\bf \overline{5}}$ under $SU(5)$.  These
fields are assumed to acquire supersymmetry breaking masses 
through  their couplings to a standard
model singlet field $X$ which develops vacuum expectation values along its
scalar as well as its $F$--components.  
Explicit models where this occurs dynamically have been constructed 
\cite{models}.  
Typically these models contain the superpotential couplings
\begin{equation}
W = \lambda_q \bar q'q'X+\lambda_l \bar\ell'\ell'X
\label{eq1}
\end{equation}
which renders a non--supersymmetric spectrum for the $(q',\ell')$ fields
through $F_X \neq 0$.  The $(q',\ell')$ fields, through their standard
model gauge interactions, induce SUSY breaking masses for the R-odd
scalars $(\widetilde{q}, \widetilde{\ell}$) and gauginos 
($\widetilde{B}, \widetilde{W},\widetilde{g})$.    
The gaugino mases arise through one--loop diagrams and are given 
(in the approximation $F_X \ll \lambda_{q,\ell} \left \langle X \right 
\rangle^2$) as
\begin{equation}
M_i(\Lambda) = {\alpha_i(\Lambda) \over 4 \pi} \Lambda~.
\label{eq2}
\end{equation}
Here $\Lambda \equiv F_X/\left \langle X \right \rangle$ is the 
effective SUSY breaking scale, which is close to the mass scale of 
the
${\bf 5}+{\bf \overline{5}}$ fields.  The scalar masses arise through
two--loop diagrams and are given as
\begin{equation}
m^2 (\Lambda) = 2 \Lambda^2\left\lbrace C_3 \left[{\alpha_3(\Lambda) \over 4 
\pi}\right]^2 
+ C_2 \left[{\alpha_2(\Lambda)\over 4 \pi}\right]^2 + { 3 \over 5} \left({Y
\over 2}\right)^2 \left[{\alpha_1(\Lambda) \over 4 \pi}\right]^2 
\right\rbrace~.
\label{eq3}
\end{equation}
We have adopted the usual $SU(5)$ normalization of the hypercharge
coupling, $\alpha_1 \equiv {5 \over 3} \alpha_Y$ and defined
$Y$ as $Q = T_3 + {Y \over 2}$.  In Eq.~(\ref{eq3}), $C_3 = 4/3$
for colored scalars and zero for sleptons, $C_2 = 3/4$ for $SU(2)$
doublets and zero for singlets.  

The mass relations of Eqs.~(\ref{eq2})--(\ref{eq3}) will
arise in a large class of models which might differ
in the details of supersymmetry breaking; it is sufficient that
SUSY breaking is conveyed to the standard model
particles through the ${\bf 5}+{\bf\overline{5}}$ fields as in 
Eq.~(\ref{eq1}).  
Thus the mass relations would hold 
even if SUSY is broken spontaneously by the 
O'Raifearteigh mechanism ($F$--type breaking).  
The ${\bf 5} + {\bf \overline{5}}$ fields is of course the simplest
extension of the standard model sector that preserves the
successful unification of the gauge couplings.  

The relations in Eqs.~(\ref{eq2})--(\ref{eq3}) imply that there is  a 
clean separation of scales among the SUSY
particles, as their masses scale with the interaction strength.
Thus the squarks are the heaviest, followed by the gluino,
left--handed sleptons $(\widetilde{\ell}_L)$, $\widetilde{W}$, right--handed 
sleptons $(\widetilde{\ell}_R$), and the
$B$--ino, which is the next-to-lightest R-odd particle
(the lightest is the gravitino $\widetilde{G}$).
Their masses (neglecting RGE-running) are in the 
ratios $11.6:7.0:2.5:2:1.1:1.0$.  
{}From an immediate experimental perspective the most
interesting particles are the lightest ones, $\widetilde{\ell}_R$, 
the chargino 
(which we shall denote by $\widetilde{W}$) and the two neutralinos
($\widetilde{N}_1$ and $\widetilde{N}_2$).  The latter are essentially 
pure gaugino
eigenstates since their mixing with the Higgsinos are suppressed by the 
large $\mu$ parameter required in the model.  

The relations in Eqs.~(\ref{eq2})--(\ref{eq3}) receive significant 
corrections from the renormalization group evolution. 
This evolution has several interesting consequences:

$\bullet$ It drives the mass-squared of $H_u$, the MSSM Higgs doublet that
couples to the top quark, which is positive
at the scale $\Lambda$, to a negative value near $M_Z$ and thus facilitates
electroweak symmetry breaking.  The mass-splitting between $H_u$ and
$H_d$ is given by
\begin{equation}
m_{H_u}^2-m_{H_d}^2 \simeq -{6h_t^2 \over 8 \pi^2} m_{\widetilde{t}}^2\, {\rm
ln}\left({\Lambda \over m_{\widetilde{t}}}\right)~.
\end{equation}
Observe that $H_u$ receives a mass correction proportional to
$m^2_{\widetilde{t}}$, which is much larger than $m^2_{H_u}$.
  
$\bullet$ It
lowers the masses of the $(\widetilde{B}, \widetilde{W})$,
and raises the masses of $\widetilde{\ell}_R$ (as well
as $\widetilde{\ell}_L$ and $\widetilde{q}$), from their bare values
at $\Lambda$.  Consequently, the mass splitting
between $\widetilde{B}$ and $\widetilde{\ell}_R$ is enhanced at the weak 
scale.  We find that a
mass splitting of $10\gev$ at $\Lambda$ typically becomes
$(20-25)\gev$ at $\mz$.  Upon including, as well,
the positive $D$--term correction
to the mass of $\widetilde{\ell}_R$, the mass splitting becomes about
$35\gev$ for $M_2\simeq200\gev$. (For definitiveness we have taken $\Lambda
=100\gev$ for the numbers given here.) 
Contours of this splitting are shown in 
Figure~\ref{fig:mass}. This difference has
significant impact on the interpretation of the observed $ee\gamma
\gamma$ event at CDF.  
\begin{figure}[t]
\centering
\epsfxsize=5in
\hspace*{0in}
\epsffile{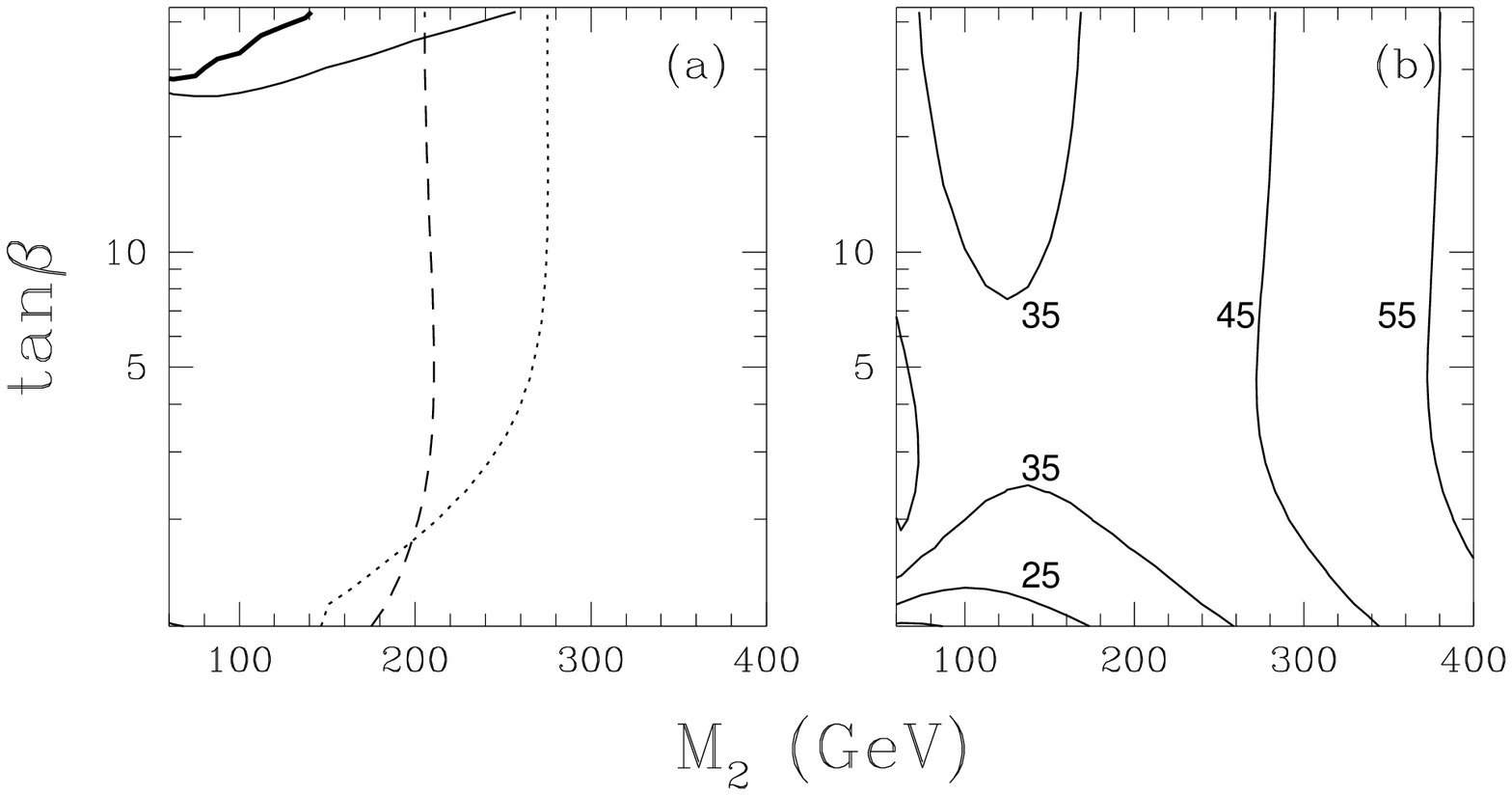}
\caption{(a) Boundaries of interesting regions in the $(M_2,\tanb)$ plane:
the region above the dark, solid line does not have a QED-preserving 
potential minimum; in the region above the light solid line, the $\staur$
is the NLSP; in the region to the right of the dashed line, the decay
$\char\to W\neut_1$ is kinematically allowed; to the right of the dotted
line the decay $\neut_2\to\neut_1 h^0$ is kinematically allowed. 
(b) Contours of 
$m_{\slepr}-m_{\neut_1}$ in GeV, for $\mu<0$.}
\label{fig:mass}
\end{figure}

$\bullet$ It supplies a sufficiently
large mass to the would--be axion of the model.  The soft
bilinear mass term $B\mu$ for the Higgs field, $(B\mu H_u H_d + H.c.)$ in the
Higgs potential, is extremely small at $\Lambda$.  If not for the effect
of running (and in the absence of other
contributions to $B(\Lambda)$), 
one would find a weak scale axion, 
in conflict with observation.  However, we find that a large enough
$B$ is generated in the process of running so that $m_A^2 = -2B\mu/{\rm
sin}2\beta \sim (400\gev)^2$.

After including the effects of renomalization group evolution and D-terms,
the mass
spectrum of the model looks as follows.  The gaugino masses obey a simple 
scaling with the the gauge couplings, so that
\begin{equation}
M_i = M_i(\Lambda) {\alpha_i \over \alpha_i(\Lambda)}~.
\end{equation}
The masses of $\widetilde{\ell}_R$,  $\widetilde{\ell}_L$ and $H_d$ are given by
\begin{eqnarray}
m^2_{\widetilde{\ell}_L,H_d} &=& m^2_{\widetilde{\ell}_L,H_d}(\Lambda) + {3 \over
2}M_2^2\left({\alpha_2^2(\Lambda) \over \alpha_2^2}-1\right) +
{1 \over 22} M_1^2 \left({\alpha_1^2(\Lambda) \over
\alpha_1^2}-1\right) \\ \nonumber
& &-\mz^2\left({1 \over 2} - s_W^2\right){\rm cos}2\beta \nonumber \\
m^2_{\widetilde{\ell}_R} &=& m^2_{\widetilde{\ell}_R}(\Lambda) 
+{2 \over 11} M_1^2
\left({\alpha_1^2(\Lambda) \over \alpha_1^2}-1\right)-\mz^2 s_W^2 {\rm
cos}2\beta 
\end{eqnarray}
where the last terms are the $D$-term contributions, and should be omitted for
$m^2_{H_d}$.
The $A$ and $B$ parameters are given by
\begin{eqnarray}
A_t\simeq A_t(\Lambda) +M_2(\Lambda)\left[-1.85+0.94 {Y_t \over Y_f}\right]
\nonumber \\
B\simeq B(\Lambda)-{1 \over 2}A_t(\Lambda)+M_2(\Lambda)
\left[-0.12+0.47{Y_t \over Y_f}\right],
\end{eqnarray}
which hold for low to intermediate values of $\tan\beta$.
Here $Y_t = h_t^2$ is the top Yukawa coupling squared, and $Y_f \simeq
2.79$.   
$A_{u,c}$ are obtained from $A_t$ by setting $Y_t=0$. The final mass ratios
for $\widetilde{q}:\widetilde{g}:\slepl:\char:\slepr:\neut_1$
are given by $9.3:6.4:2.6:1.9:1.35:1.0$.  (All of our numerical results
are derived using the complete one--loop RGE's.)

In the minimal messenger 
model without any modification of the Higgs sector, one
expects $A$ and $B$ parameters to be negligible at $\Lambda$,
since they arise only from higher loops (see however Ref.~\cite{dvali}).  
However, in the process of
running, significant $A$ and $B$ can be induced. Furthermore, in this
constrained version, the parameter tan$\beta$ and the sign of $\mu$
are determined.  These results 
follow from minimizing the Higgs potential,
which leads to 
\begin{eqnarray}{\rm sin}2\beta &=& {-2B\mu \over m_{H_u}^2+m_{H_d}^2+2\mu^2}
 \nonumber \\
\mu^2 &=& {m_{H_d}^2 - m_{H_u}^2{\rm tan}^2\beta \over {\rm
tan}^2\beta-1} - {1 \over 2}\mz^2~.
\label{eq9}
\end{eqnarray}
The last relation implies that $\mu$ is large in the model.  If
$B(\Lambda)=0$, we can infer its effective value at $\mz$.  
Tan$\beta$ is then determined through the RGE-improved tree-level potential
of Eq.~(\ref{eq9}) to lie at $\tan\beta\approx
30$; estimates of the full 1-loop contributions appear to push $\tan\beta$
up to values near $\tan\beta\approx 40$ to 50.
In this most economical version, one also finds
the sign of $\mu$ to be negative.  

A further constraint arises from demanding that the $\staur$
mass--squared not turn negative (otherwise the vacuum will be superconducting).
In Figure~\ref{fig:mass}\ we show the contour
of $m_{\staur}^2=0$ in the tan$\beta-M_2$ plane;  $\tan\beta \gsim 50$
is excluded.  A more stringent constraint along these lines
is to require, in the spirit of the phenomenology we are attempting to
address, that 
$m_{\widetilde{\tau}_R} > m_{\widetilde{N}_1}$,
otherwise $\widetilde{N}_1$ will decay
dominantly into $\tau+\widetilde{\tau}_R$, destroying the  $\gamma \gamma$
signature.  The contour where $\widetilde{\tau}_R $ mass equals that of
$\widetilde{N}_1$ is also shown in Figure~\ref{fig:mass}; above that line
the $\staur$ is the lighter. Because the bound corresponds roughly to 
$\tan\beta \sim 25$, we shall 
focus only on the range $\tan\beta \le 25$ in 
presenting our numerical results.


\section{Production and Decay of Sparticles}

Given the physical parameters $M_2$, $\tanb$ and $\sgnmu$ 
all cross-sections and
branching ratios within the 
MMM can be calculated. It is particularly important that
within the MMM 
there are three widely separated mass scales:
light particles with only $U(1)$ hypercharge interactions, intermediate
mass particles with $SU(2)_L$ interactions, and heavy particles with
$SU(3)_C$ interactions. The heavy states (squarks and gluinos) are typically
around $1\tev$ for light particle masses consistent with LEP bounds;
therefore their production rates even at the upgraded FNAL will be orders
of magnitude below observable signal levels. Finally, the electroweak
breaking constraint usually fixes $\mu$ such that $M_2\ll|\mu|\lsim M_3$,
pushing sparticles whose masses come from $\mu$ (\ie, higgsinos and the
second doublet of Higgs bosons) out of the reach of current accelerators.


On the other hand, the light and intermediate mass sparticles can have
non-negligible production rates even in Run I of FNAL ($\int{\cal L}
\simeq 120\invpb$). For this study of the MMM, we consider a
$p\bar p$ collider/detector with $\sqrt{s}=1.8\tev$, as at FNAL Run I, with
100\% efficiency for tagging all relevant final states. Because the sparticles
are always produced in pairs, and because the gluinos and squarks are too
heavy to be produced at reasonable $x$, we need only consider Drell-Yan (DY)
production processes. Each process is calculated at the parton level
for $\hat s=x_1x_2s$, then
integrated over the proton parton distribution functions, $f_{q/p}(x_i)$,
in this case using those of the CTEQ3M set \cite{cteq}. Previous calculations
of DY sparticle production were used or altered for the parton level 
processes studied here~\cite{ehlq}. The parameter
space studied is spanned by the three parameters $(M_2,\tanb,\sgnmu)$ in
the ranges $50\le M_2\le 400$, $1<\tanb\le25$ and $\mu$ taking both
signs.

There are a limited number of sparticle DY processes which can be expected
to be observable at the current generations of hadron colliders:
$$
p\bar p\to Z\to\left\lbrace\begin{array}{c}\sneut\sneut^* \\
\slepl\slepl^* \\
\slepr\slepr^* \\ \char^+\char^- \\ \neut_i\neut_j \end{array}
\right. \quad\quad 
p\bar p\to W\to\left\lbrace\begin{array}{c}\slepl\sneut^* \\
\char\neut_i \end{array}\right.
$$
where $i,j=1,2$. Not surprisingly, in order for production rates of the
heavier slepton modes ($\sneut\sneut$, $\slepl\slepl$) to be observable, 
rates for the lighter sparticles would be much larger, already providing an
unambiguous signal at a collider. Since  a large signal has
in fact not been observed we will not consider the DY production of
left-handed sleptons further.

One also expects (and we have explicitly verified) that the rate
for double neutralino production ($\neut_i\neut_j$) is small
compared to the others, since in the limit $\neut_1=\bino^0$ and $\neut_2=
\wino^0$ there are no $Z\neut_i\neut_j$ interaction terms in the MSSM.
(This fact also implies that the calculation of $\neut_1$ mass bounds
from non-observation of $Z\to\neut_1\neut_1$ at LEP1.5 need not
coincide with the kinematic limit of approximately $65\gev$; in fact
we find the appropriate mass bound from OPAL \cite{opal}\ to be
$10-15\gev$ below this limit.) Likewise, the $\neut_1\char$ channel
is suppressed due to the lack of a $W^+\wino^-\bino$ interaction. We
find the production rate of $\neut_1\char$ to be at least two orders of
magnitude below the rates for competing processes, making it inaccessible
at current colliders.

The dominant DY production channels are then $\slepr\slepr$,
$\neut_2\char$, and $\char\char$. In Figure~\ref{fig:cross}
we have shown contours of cross-sections for the three production channels
in the $(M_2,\tanb)$ plane, where we have summed over slepton flavors and
gaugino charges. The contours are labelled in femtobarns (fb),
and one should recall that $\sigma=10\fb$ corresponds to 1 event in
$100\invpb$ of data. One notes some general features: {\sl (i)} There is
little or no dependence on $\sgnmu$ in the most important channels, the largest
dependence on $\sgnmu$ coming in the $\char\neut_1$ channel, which is several
orders of magnitude below those shown; {\sl (ii)} the cross-sections are
only weakly dependent on $\tan\beta$; {\sl (iii)} at large $M_2$, the slepton
production is largest, while it is the gaugino production which dominates at 
small $M_2$; in particular, when the cross-sections are near $10\fb$, all 
three cross-sections are within a factor of four of each other; {\sl (iv)}
for $M_2\lsim 160\gev$, the production rate for gauginos would be very
large even now at the Tevatron.

\begin{figure}[t]
\centering
\epsfxsize=3in
\hspace*{0in}
\epsffile{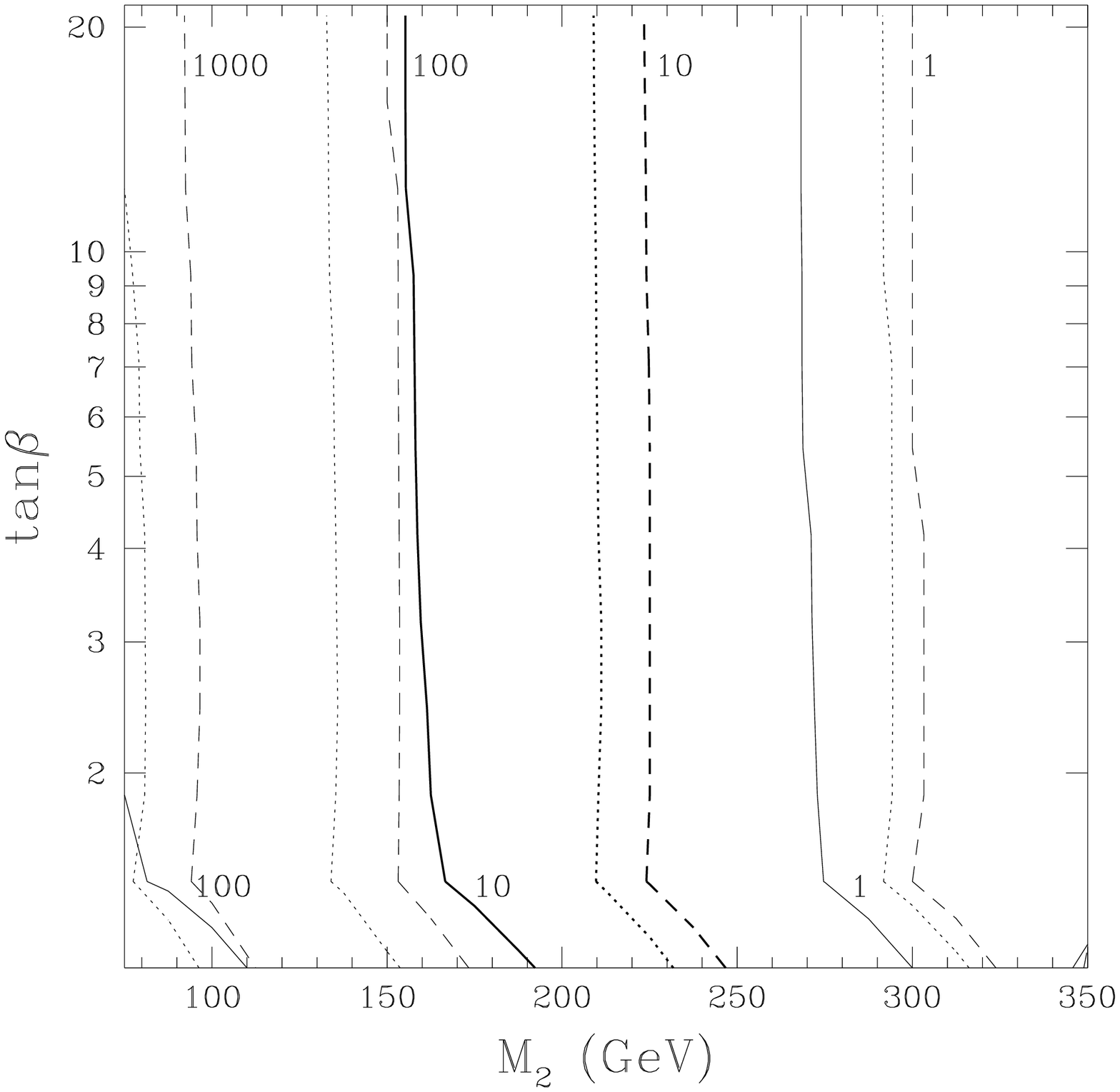}
\caption{Cross-sections at the Tevatron for $\slepr\slepr$ (solid),
$\char\neut_2$ (dashed), and $\char\char$ (dotted) production.
(Slepton flavors and gaugino charges summed over.)
The contours are labelled in fb; the labels at the top of the plot label
both gaugino production channels, while those at the bottom label the
slepton production. The contours
for 1 event in $100\invpb$ are in bold.}
\label{fig:cross}
\end{figure}

The exact signature of each channel will depend on how its constituents
decay. Consistent with the MMM and the CDF event, 
we take the gravitino to be the lightest R-odd particle, 
to which the next-lightest R-odd particle (the $\neut_1$)
decays  via $\neut_1\to\gamma\gravitino$.  Then essentially all
sparticles produced will eventually decay to $\neut_1$ and in turn
to $\gamma+\misset$; thus $\gamma\gamma+\misset$ 
is a mandatory ingredient of the final state in
all MMM sparticle production processes. 

Within the MMM, the $\slepr=\widetilde e_R,\widetilde\mu_R$ and the 
$\widetilde\tau_R$
can only decay directly to the 2-body final state $\ell\neut_1$ or 
$\tau\neut_1$, where 
$\ell=e,\mu$ (we treat $\tau$ leptons separately for later convenience).
Since the production cross-section for each flavor of slepton is the
same, one observes a signal $\tau^+\tau^-\gamma\gamma+\misset$ 1/3 of the
time, and $\ell^+\ell^-\gamma\gamma+\misset$ 2/3 of the time. Such a 
final state describes the $ee\gamma\gamma+\misset$ event at CDF.

There has been some discussion in the literature as to whether the MMM could
describe the CDF event through $\slepr\slepr$ production given the small mass 
splitting between $\slepr$ and $\neut_1$~\cite{dtw}. The authors of 
Ref.~\cite{dtw}\ were concerned because one of the electrons in the CDF event
is highly energetic ($E_\perp \simeq 59\gev$) while they assumed 
$m_{\slepr}\simeq 
1.1\,m_{\neut_1}$, which in the mass range of interest corresponds to
$\slepr-\neut_1$ mass splitting of only about $10\gev$.
The required boost for the initial pairs of sleptons seemed to reduce 
significantly the probability of interpreting this event as coming from MMM
$\slepr$ production. This may be too pessimistic given the
analysis of the $\slepr$--$\neut_1$ mass difference in the previous Section.
{}From that analysis, we find for the appropriate mass range that 
$m_{\slepr}-m_{\neut_1}\simeq35\gev$ and the required boost is small ($\gamma
\simeq 1.3$). However, as pointed out in~\cite{dtw}, larger mass differences
in turn imply smaller rates for $\slepr\slepr$ production with
respect to the $\char\neut_2$ and $\char\char$ channels. 
For a $35\gev$ mass splitting, this suppression is about a factor of two 
relative to a $10\gev$ splitting. Given the current
paucity of data, this is completely consistent with observation.

Because $\neut_2$ and $\char$ are nearly degenerate in mass, the 
preferred decay of the heavier of the two to the lighter via 
$\char^\pm\to\neut_2 W^\pm$, cannot occur. If $m_{\char}-m_{\neut_1}
>\mw$ then the 2-body decay $\char^\pm\to\neut_1 W^\pm$ will be allowed.
In Figure~\ref{fig:mass}(a) the contour along which the decay becomes
kinematically allowed is displayed.
However, the coupling is suppressed by the lack of any $W^\pm\wino^\mp\bino$
interaction in the MSSM, so unsuppressed 3-body decays can compete with the
2-body mode. (Calculations 
of all relevant 2- and 3-body decay widths exist in the 
literature~\cite{decays}.)
The 3-body decays proceed through t-channel exchange of
$\slepl$ or $\sneut$: $\char\to\ell\nu_\ell\neut_1$. (In this way it
might be 
possible to account for the CDF event 
using $\char\char$ production, 
without
having to invoke coincidence against
a $\frac{1}{9}\times\frac{1}{9}$ suppression in the $W$
decay branching ratios.  A signal for this possibility 
is the existence of 
equal numbers of like and unlike lepton pair events, 
\eg, both $e^+e^-\gamma\gamma+\misset$ and $e^+\mu^-\gamma\gamma+\misset$.)

At large $\tanb$ there is a strong enhancement in the 3-body decay into
$\tau\nu_\tau\neut_1$. This arises because at large $\tanb$ the $\tau$-Yukawa
coupling becomes $\ord(1)$ and couples to the $\higgsino_D$ component
of $\char$, which can be relatively large. For example, for
$\mu$ positive (negative)
and $\tanb$ greater than 5 (10), the branching ratio of $\char\to
\tau\nu_\tau\neut_1$ exceeds 50\%.

The decay channels of $\neut_2$ are similarly a mix of 2-body decays
suppressed by small couplings and 3-body decays suppressed by phase space.
The three 2-body decays are to $\neut_1h^0$, $\neut_1Z$ and $\slepr^\pm
\ell^\mp$ final states. The first two can be either allowed or
disllowed in alternative  
large regions of the parameter space; the third is always allowed. Each
2-body process suffers from small couplings which go to zero in the limit
$\neut_2=\wino^0$ and $\neut_1=\bino^0$. For small gaugino-higgsino mixing
the leading 2-body coupling is through the $\tau$-Yukawa coupling so that 
even at relatively small $\tan\beta$, decays to $\tau$'s predominate.

When it is kinematically allowed,
the decay $\neut_2\to\neut_1h^0$ has a large ($\sim 50\%$) branching fraction.
Such a process could provide an interesting route toward pursuing Higgs
physics at the Tevatron, with Higgs production neatly tagged by
the hard photon pair. 
Because the lightest Higgs boson, $h^0$, receives large radiative corrections
to its mass, there is considerable uncertainty about the precise prediction of
$m_{h^0}$ for each point in our parameter space. Further, 
$m_{h^0}$ is sensitive 
to deviations from {\sl minimal} messenger models, for
example to inclusion of a singlet Higgs field 
(perhaps useful for generating a $\mu$-term).
Therefore
in presenting our results, we have pretended that the Higgs decay mode is
disallowed; if it is allowed at all it is appreciable.
We will highlight where applicable the swath through parameter
space where the Higgs mode is open in the truly-minimal MMM.
When on-shell Higgs production is disallowed, the coupling suppression serves
to make any 3-body decay through the Higgs unobservably small. The contour
along which the Higgs decay mode becomes accessible is shown in 
Figure~\ref{fig:mass}(a).

The two 3-body decays of $\neut_2$ are via t-channel $\slepl$ and $\sneut$:
$\neut_2\to\ell^+\ell^-\neut_1$ and $\neut_2\to\nu\bar\nu\neut_1$. 
At large $\tanb$ the $\tau\tau\neut_1$ final state receives 
substantial enhancement from the $\higgsino_D$ component of $\neut_2$ coupling
to $\staul\tau$ via the large $\tau$-Yukawa coupling, as discussed in the
case of $\char$ above. This means that for large $\tanb$, the production
of $\char^+\neut_2$ will give a signal of $\tau^+\tau^+\tau^-\gamma
\gamma+\misset$ nearly 100\% of the time, while $\char\char$ production
will look the same with one fewer final state $\tau$-lepton.

In Figure~\ref{fig:sigmabr}\ we have folded the branching ratios into the
production cross-sections to show the expected production rates for various
sets of ``final'' states in the detector. Here we have defined on-shell
$W^\pm$, $Z$, and $h^0$ to be final states; the reader can identify the
observed final states associated with production of these particles simply
by multiplying by their well-known branching ratios. By keeping the gauge
bosons in the final state, one also learns about the invariant 
masses of their decays products. Specifically, 
Figure~\ref{fig:sigmabr}\ displays the $\sigma\cdot$BR of nine final states for
$M_2=200\gev$ and varying $\tanb$. The two pairs of plots show the
results for both signs of $\mu$.

Figure~\ref{fig:sigmabr}(a) shows the five most relevant channels at 
small $\tanb$ and $\mu>0$ 
(all channels that follow include $\gamma\gamma+\misset$): 
$\ell\ell$ (solid), $W\ell\ell$ (dotted), $W\tau\tau$ (short-dashed),
$WW$ (long-dashed), and $WZ$ (dot-dashed). Figure~\ref{fig:sigmabr}(c)
is the same for $\mu<0$. Figure~\ref{fig:sigmabr}(b) shows the four most
relevant channels at large $\tanb$ and $\mu>0$ (not including $\ell\ell$):
$\tau\tau\tau$ (solid), $\tau\tau$ (dotted), $\tau\ell\ell$ (short-dashed),
and $W\tau$ (long-dashed). Figure~\ref{fig:sigmabr}(d)
shows the same five curves for $\mu<0$. In all four plots, the shaded region
is the portion of parameter space that would allow the decay $\neut_2\to
\neut_1h^0$.
\begin{figure}
\centering
\epsfxsize=5in
\hspace*{0in}
\epsffile{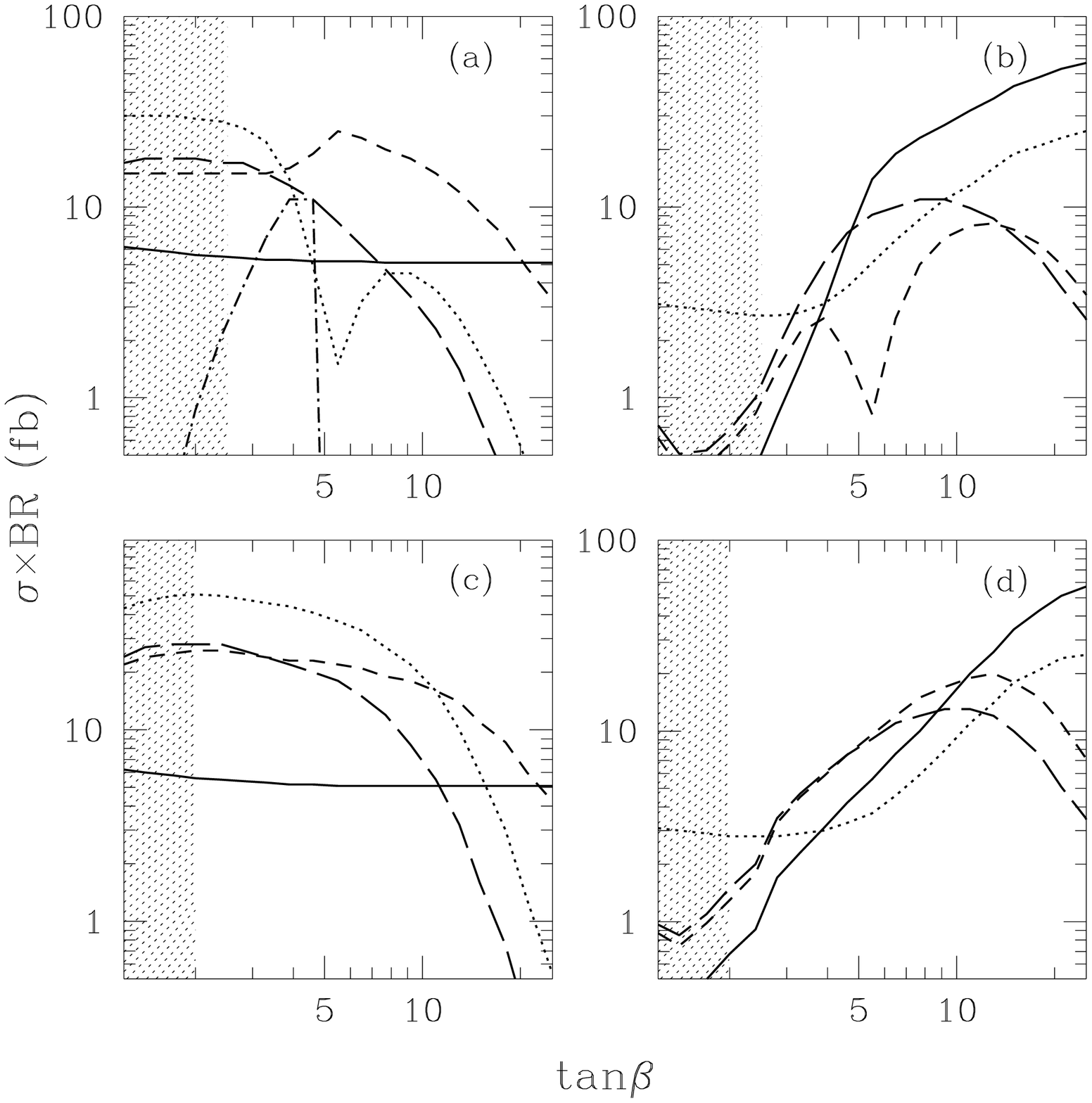}
\caption{Cross-section times branching ratio for $M_2=200\gev$ with $\tanb$
varied along the x-axis. Both signs of $\mu$ are shown: (a)--(b) $\mu>0$,
(c)--(d) $\mu<0$. The nine most relevant channels are displayed, 
divided between
two subfigures for each $\sgnmu$. The shading is in the region for which
$\neut_2\to\neut_1h^0$ is allowed in the minimal model.
See the text for a full description.}
\label{fig:sigmabr}
\end{figure}

{}From this analysis, it
clearly emerges that if one is to interpret the CDF event to suggest
$\sigma\times BR(p\bar p\to \slepr\slepr)\simeq(5-10)\fb$, 
then the MMM implies that there are 
several other channels which should also be accessible at the Tevatron. 
If $\tanb<5$ or 10 (depending on $\sgnmu$), then the largest available channel
is $W\ell\ell$ (for $\ell=e,\mu$) with $W$ decaying into 2 jets (2/3 of the
time) or a single lepton (1/3 of the time): $p\bar p\to\ell^+\ell^-jj\gamma
\gamma+\misset$ or $p\bar p\to\ell^+\ell^-\ell^\pm\gamma\gamma+\misset$. Second
to this rate will be $WW\gamma\gamma+\misset$ channels (with the usual $W$
decay branching ratios) and $W\tau\tau\gamma\gamma+\misset$. There is
also a small window around $3\lsim\tanb\lsim5$ and $\mu>0$ in which 
$WZ\gamma\gamma+\misset$ is an important final state.

For large $\tanb$, $\tau$ production of all kinds dominates. If $\tanb\gsim
5(10)$, then the Tevatron should be producing $(n\tau)X\gamma\gamma+\misset$ at
rates comparable (or larger than) $ee\gamma\gamma+\misset$, for $n=1,2,3$.


\section{Discussion}

We have described a minimal implementation of a low-energy messenger
model for supersymmetry breaking.  Outstanding features of the model
include: generic production of final states including $\gamma \gamma$
and missing energy; 
clean separation of scales in the R-parity odd spectrum, with the
squarks and gluinos much heavier than the electroweak 
gauginos and sleptons; and
very small mixing of the $SU(2)\times U(1)$ neutralinos with Higgsinos
and with each other.  Many
specific features are predicted in terms of very few parameters, so
that the model could be rigorously tested, or eliminated, with quite a
small amount of relevant data.  If the CDF 
$e^+e^-\gamma \gamma$ event represents a first indication, 
such data is not far beyond current reach.

Of course in insisting upon the absolutely minimal model we may have
overreached.  We have shown that some apparent difficulties with the
minimal model are removed upon careful treatment of coupling
renormalization and proper inclusion of D-terms. 
Nevertheless, one must remain open to such possibilities as different
overall mass scales for the sleptons as against the gauginos, or
additional contributions to the parameters of the Higgs sector.   

Finally let us briefly mention another possibility for physics capable of 
generating the CDF $e^+e^-\gamma \gamma$ event, which if nothing else
might serve as a useful foil.  If there were heavy pion-like
particles whose decay modes were dominantly 
$\Pi^\pm \rightarrow W + \gamma$ and 
$\Pi^0 \rightarrow ZZ ~{\rm or}~ Z\gamma ~{\rm or}~ \gamma
\gamma$, their production and decay could induce many of the same
final states as we have analyzed above, including the CDF event.  The
missing energy in this event would be carried off by neutrinos; but in
other cases one might have $\gamma \gamma $ without significant
missing energy, so that this alternative will be readily
distinguishable.

\section*{Acknowledgements}
We would like to thank J.~Conway, H.~Frisch, J.~March-Russell and S.~Treiman 
for many stimulating discussions.


\end{document}